\documentclass[runningheads]{llncs}
\usepackage{graphicx}
\usepackage{hyperref}
\usepackage{multirow}
\usepackage{etoolbox}
\usepackage{float}
\usepackage{ragged2e}
\usepackage[many]{tcolorbox}
\usepackage{cleveref}
\usepackage{xcolor}
\usepackage{parskip}
\usepackage{amssymb}
\usepackage{ifmtarg}
\usepackage{subcaption}
\usepackage{svg}
\usepackage{tabularx}

\newcommand\ainote[1]{\textcolor{red}{[AI: #1]}}
\newcommand\jnote[1]{\textcolor{blue}{[J: #1]}}
\newcommand\cnote[1]{\textcolor{magenta}{[C: #1]}}
\newcommand\knote[1]{\textcolor{green}{[K: #1]}}
\newcommand\fnote[1]{\textcolor{violet}{[F: #1]}}

\newcommand\ignore[1]{}

\newenvironment{tweet}{%
  \newcommand{\tweetItself}[2]{\flushright ##2}%
  \begin{tcolorbox}[size=small,nobeforeafter]%
  \RaggedRight%
}{\end{tcolorbox}}

\begin{document}
%
% \title{Clustering Coordinated Information Campaigns on Social Media with Cross-Platform and Multi-Modal Embeddings}
\title{Multi-Modal Embeddings for Isolating Cross-Platform Coordinated Information Campaigns on Social Media}
\titlerunning{Cross-Platform Coordinated Campaigns}
\author{
Fabio Barbero \and
Sander op den Camp \and Kristian van Kuijk \orcidID{0009-0008-3013-8760} \and Carlos Soto García-Delgado \and Gerasimos Spanakis\orcidID{0000-0002-0799-0241} \and Adriana Iamnitchi\orcidID{0000-0002-2397-8963}
}
\authorrunning{F. Barbero et al.} 
\institute{Department of Advanced Computing Sciences\\ Maastricht University, The Netherlands\\
\email{\{fabio.barbero,sjgm.opdencamp,kristian.vankuijk,\\c.sotogarciadelgado\}@student.maastrichtuniversity.nl}\\\email{\{jerry.spanakis,a.iamnitchi\}@maastrichtuniversity.nl}}
\maketitle              % typeset the header of the contribution
\begin{abstract}
Coordinated multi-platform information operations are implemented in a variety of contexts on social media, including state-run disinformation campaigns, marketing strategies, and social activism.
Characterized by the promotion of messages via multi-platform coordination, in which multiple user accounts, within a short time, post content advancing a shared informational agenda on multiple platforms, they contribute to an already confusing and manipulated information ecosystem. 
To make things worse, reliable datasets that contain ''ground truth`` information about such operations are virtually nonexistent. 
This paper presents a multi-modal approach that identifies the social media messages potentially engaged in a coordinated information campaign across multiple platforms. 
Our approach incorporates textual content, temporal information and the underlying network of user and messages posted to identify groups of messages with unusual coordination patterns across multiple social media platforms. 
We apply our approach to content posted on four platforms related to the Syrian Civil Defence organization known as the White Helmets: Twitter, Facebook, Reddit, and YouTube. 
Results show that our approach identifies social media posts that link to news YouTube channels with similar factuality score, which is often an indication of coordinated operations. 
%We discuss our results and the limitations of our approach. 
%Our evaluation strongly hints towards coordinated behaviour within our clusters. 
%This research contributes to the development of robust methods for detecting and analysing information campaigns on social media, aiding in the fight against information and promoting a more informed public discourse.

\end{abstract}

\section{Introduction}

While social media platforms may be independently attempting to detect nefarious information campaigns within their own platforms, the problems due to multi-platform coordination are not likely to be addressed in the current environment of legislature void and platforms' competition for user attention. 

Previous research has shown that information operations online are now much more sophisticated than networks of coordinated bots that (re)broadcast some messages. 
Instead, such coordinated efforts 
i)~are concurrently deployed on multiple social and alternative news media platforms~\cite{ng_multi-platform_2021,ganguly_aims_2023}; 
ii)~promote the intended message by means specific to each platform (such as videos on YouTube and written news or opinion articles on alternative media websites)~\cite{wilson_cross-platform_2021,thomson_strategic_2020}; 
iii)~amplify the same message by repeating it in altered forms (such as including the same video footage in multiple videos posted on different channels on YouTube)~\cite{ng_multi-platform_2021,thomson_strategic_2020}; 
and iv)~amplify the same content via platform-specific affordabilities, such as retweeting on Twitter vs. liking/commenting on YouTube to possibly manipulate the platform's content promotion algorithms~\cite{lukito_coordinating_2020,thomson_strategic_2020}. 

%,\cite{kellegupta2019malreg,keller2020political}

Due to limited datasets on such covert, often undetected operations, previous research focused on specific contexts (for example, the Russian Internet Research Agency's engagement in influencing the US 2016 elections) and on single platforms, typically the promotion of specific URLs on Twitter. %~\cite{graphika's paper} 
Yet coordinated online information campaigns continue to be employed by various state, business, or social actors for various purposes, from disinformation and monetization to activism and public health campaigns.
Although some are carried out by legitimate groups or individuals, many campaigns may be supported by entities with undisclosed motivations, such as foreign governments or political parties, and may not be transparent about their underlying goal~\cite{wilson_cross-platform_2021,lukito_coordinating_2020,nghiem_detecting_2021}.
While social media platforms reacted to some of such problems and typically blocked accounts seen to engage in coordinated information operations~\cite{noauthor_moderation_nodate}, coordinated operations deployed on multiple platforms can escape the vigilance of individual platforms if they keep a low profile on any particular platform.

This paper proposes a methodology that, given a multi-platform collection of social media posts, identifies a subset of such posts that are likely to be part of a coordinated information campaign and clusters them based on content and posting time locality. Clusters with only one post are likely to represent messages that do not participate in a coordinated campaign, thus ignoring them can reduce the dataset that needs to be further analysed by human operators or via computational means. We show on a multi-platform datasets that includes content posted on Twitter, Facebook, Reddit and YouTube that by using multi-modal approaches (text, network structures, and time) we can identify similar messages that not only link to the same URL (as previous work has been done) but link to different URLs of similar factuality as reported by the Media Bias/Fact Check website\footnote{https://mediabiasfactcheck.com/}. In the lack of ground-truth datasets that clearly mark which messages are part of a coordinated operation, we use agreement on factuality within a cluster as a performance metric. 

Our approach can work on diverse combinations of platforms and on diverse topics without the requirement to identify whether messages are factual or not. This generality is due to the design of our approach that: 1)~considers only messages that introduce new information on the platform (e.g., tweets on Twitter but not retweets; posts and comments on Reddit, etc.), thus not depending on platform-specific message promotion algorithms or platform-specific functionalities; 2)~it detects coordination in the absence of shared tokens (such as hashtags or URLs). However, our technique works under the assumptions that coordinated campaigns push syntactically similar (but not identical) messages within a short interval of time.

\section{Related Work}
\label{sec:related_work}
Intentional spread of fake news with political or financial objectives has been studied for cases such as Brexit\cite{calisir_Twitter_2022}, COVID-19~\cite{chen_tracking_2020}~\cite{deverna_covaxxy_2021} and the Syrian War~\cite{abu_salem_fa-kes_2019}. Sometimes, these fake news are known to be posted and promoted by bots~\cite{ganguly_aims_2023}. The behaviour of these bots has been studied, and different datasets exist for their activity on platforms such as Twitter~\cite{davis_botornot_2016}~\cite{shao_spread_2018}.

Furthermore, such campaigns are also known to often be coordinated. 
One of the first studies was performed by  Ratkiewicz et al.~\cite{ratkiewicz_detecting_2011}. They describe a machine learning framework that combines topological, content-based and crowdsourced features of information diffusion networks to detect the early stages of viral spreading of political misinformation.
Sharma et al.~\cite{sharma_identifying_2020} investigates user activity embeddings to find coordination between accounts and is trained using labelled data. 
Because labelled data are not always available, others have looked at unsupervised clustering methods. 
Fazil and Abulasih~\cite{fazil_socialbots_2020}, Pachero et al.~\cite{pacheco_uncovering_2021} and Magelinski et al.~\cite{magelinski_synchronized_2021} all explored creating a user network based on similarities in user profiles, activity, or post content. 
They cluster the resulting network to identify groups of users showing signs of coordination. 
These methods have each focused on one category of features but have not looked much at combining different modalities. 
In addition, all of them use only single-platform data collected from Twitter.% , one of the largest social media platforms. 
% maybe include more limitations of existing research on clustering users

However, there is significant evidence to suggest that these coordinated campaigns do not limit themselves to only one social media platform. 
Lukito et al.~\cite{lukito_coordinating_2020} explore the activity of the Internet Research Agency (IRA), a Russian company engaged in online propaganda and influence operations, on three social media platforms, Facebook, Twitter, and Reddit, to understand how the activities on these sites were temporally coordinated. 
NG et al.~\cite{ng_multi-platform_2021} studied how coordinated campaigns occurred for the White Helmets across YouTube, Facebook and Twitter, and how they can exhibit certain patterns when studying time and relationships between users and posts. 
Our paper uses multi-modal data (text, network structures and time) to identify likely coordinated groups that post similar content in a synchronized manner on multiple platforms. % data related to videos allegedly about the White Helmets
\section{Datasets}\label{sec:datasets}
%\ainote{points to make: WH campaign was run on YouTube primarly, as it was shown before. That's why we start with WH videos to be able to select the messages from Reddit, Twitter and Facebook. We do not consider the messages on YouTube in our dataset -- need to figure out how to justify this. We start from dataset previously used in~\cite{ng_multi-platform_2021}, to which we add reddit data and factuality scores to some YouTube channels. Data collections -- briefly on all, as nothing is special. Data characteristics -- refer to table. Might not need subsections}

One topic known to have been part of information campaigns is the White Helmets~\cite{ng_multi-platform_2021}.
The White Helmets (WH)~\cite{noauthor_support_nodate} is a volunteer organization that operates in rebel held areas of Syria. They perform medical evacuation, urban search and rescue in response to bombing, evacuation of civilians from danger areas and often publish videos showing the human impacts of the conflict. The WH is known to have been a target of information campaigns by supporters of the Syrian president Bashar al-Assad and Russian state-sponsored media organizations such as Russia Today (RT) and Sputnik, for instance, accusing the WH of faking evidence of atrocities or organ-harvesting racket~\cite{noauthor_syria_2020}.

\ignore{
infeasible. In particular, while Twitter provides a dataset available to researchers on information campaigns through the \textit{Twitter Moderation Research Consortium}~\cite{noauthor_moderation_nodate}, we have found that a significant proportion of these tweets appear to contain content related to other types of spam. Furthermore, there is no easy way to get "ground truth" data that is representative of the whole of a social media platform. The language and content of tweets and other social media posts vary greatly over time and across regions, and taking a subset of such data is unlikely to be representative. Finally, most social media platforms, such as Twitter, Facebook and YouTube, take active steps to remove content that violates their policies, so much of the historical data on identified campaigns is unavailable. In the case of Twitter, no publicly available dataset can include the (non-randomised) tweet ID, meaning that a great deal of trust must be placed in the data collector. Access to APIs such as Twitter and Facebook can also be complicated, with long waiting times and hidden behind a paywall. Finally, most social media sites do not provide access to the full metadata of their posts, such as who liked the post and when. This further limits the type of information that can be used to build models.
}

For this paper, we used data from the White Helmet campaign as described in~\cite{ng_multi-platform_2021}, which we expanded with our collection of Reddit data and augmented with factuality scores from Media Bias/Fact Check. 
This dataset was originally built by collecting YouTube videos that include "White Helmets" in different languages in the video's title or description. 
The videos identified this way were posted between September 2006 and April 2019. 
Each video is uniquely identified by a \textit{video ID} and contains information on the \textit{video title, video channel name, published time} and \textit{video captions} (which can be automatically generated by YouTube). 
Comments on the videos are not included, nor the number of users subscribed to the channel.
Using the GNIP Twitter API in early 2019, tweets that include the YouTube videos URLs on White Helmets in our collection were collected. 
Later on, data was collected from CrowdTangle on Facebook messages posted between April 2018 and April 2019 that include the YouTube video URLs from the original dataset. 

\ignore{
Initial data was provided to researchers from~\cite{ng_multi-platform_2021} by an external party, and only contained Twitter posts together with some anonymized YouTube IDs. The anonymized YouTube IDs were linked to the rest of the information of the videos. Authors from~\cite{ng_multi-platform_2021} then queried the YouTube API with the video titles, and tried to retrieve the YouTube ID. While this does not seem to correctly work in all cases, more than $80\%$ of videos have a publishing time matching with the one in the original dataset.
The same researchers then used the Facebook API to retrieve more posts linking to the YouTube videos in the dataset.
}

% \ainote{not sure we should keep this. transparency vs. too much info?} \knote{i would remove it}
% We observed some inconsistencies in the dataset, such as $0.75\%$ of posts having a publishing data prior to the publishing date of the video they linked to. 
% As we were not shared the exact script used for data collection (since multiple parties were also involved), we decided to continue our research, as this means our research computes a lower bound.

%\subsection{Reddit data collection}
%\label{sec:Reddit_data}
%While the current dataset from Wai NG et al.~\cite{ng_multi-platform_2021} already contains data from multiple platforms, we augmented this data by adding posts from Reddit. 
To collect Reddit data, we performed an identical platform search to NG et al~\cite{ng_multi-platform_2021} by fetching posts during the same time frame (April 1, 2018 to April 30, 2019) containing videos from our dataset. 
We used the Pushshift Reddit Dataset~\cite{baumgartner_pushshift_2020} in March 2023. 
Pushshift has collected Reddit data and made it available to researchers since 2015, updated in real-time, and including historical data back to Reddit’s inception~\cite{baumgartner_pushshift_2020}. 
%\ainote{trouble is that they stopped providing this data not long after you collected it. We need to ack briefly this.}\knote{See the end of the paragraph. Hopefully that's enough? We don't know really know more also because Sander download it the data from a torrent.} 
Pushshift includes submissions (posts) and comments that have since been deleted by Reddit or subreddit moderators. Because of the free availability of this data, we also do not have to limit our search to specific keywords, but can find every submission and comment made that includes any of the relevant URLs from the collected YouTube data. Nevertheless, shortly after our data collection, Reddit officially removed their API access. While Pushshift still retrieves Reddit data, access is only granted for moderation purposes to approved Reddit moderators.

Each message in our dataset is identified by a unique \textit{post ID}, and has information on the \textit{user ID, published time, platform, text content, action type} (whether it is a post or a reply) and linked \textit{video ID}.
Through a language detection model, we noticed that more than $83\%$ of the posts are in English, with the second most used language being Arabic with $6.9\%$ of posts. 
We present the characteristics of our dataset in Table~\ref{tab:dataset}.

%\subsection{Augmenting Data with Factuality Scores}
%\label{sec:MBFC}
Our cross-platform dataset thus contains posts from Twitter, Facebook and Reddit that include URLs to videos that have "White Helmets" in their descriptions or titles on YouTube. 
This dataset may include authentic messages in addition to messages that may be part of the well-documented coordinated campaign against the White Helmets. 
Since the dataset is unlabelled, we augmented it with the Factuality Score on the Media Bias/Fact Check (MBFC) website~\cite{noauthor_media_2023}. 
The factuality score is an integer ranging from $0$ to $5$, with $0$ being ``Very Low'' and $5$ ``Very High'' that is assigned currently to more than $6,800$ news websites based on information from fact-checkers that are either a signatory of the International Fact-Checking Network or have been verified as credible by MBFC. 

We labelled $\approx 30$ questionable video channels and $\approx 30$ mostly factual video channels, with questionable sources being the most frequent in the dataset. 
In fact, the three most posted video channels have a very low factuality score. 
Clarity of Signal, RT and Vanessa Beeley are present in $2505$ posts, $2073$ and $1824$ posts, respectively. 
A total of $9757$ posts in our dataset posted a video labelled with a factuality score, with $7273$ posts having a factuality score of $0$ (74.5\%) and $1888$ posts a factuality score of $1$ ($19.4\%$).
%\knote{I changed to the three most posted video channels have a very low factuality score, as I labelled vanessa beeley as very low factuality. Orinally, she is not in MBFC hence why she was the unlabelled channel.}

\begin{table}[htbp]
    \centering
    \caption{Characteristics of the multi-platform dataset on White Helmets. YouTube video URLs are used to collect relevant messages posted on the other social media platforms during the April 2018--April 2019 interval.} 
    
    \begin{tabular}{l|r|r|r|r}
            \hline
        Platform & Posts & YT URLs & YT channels & Users\\
        \hline
        Reddit & $481$ & $113$ & $63$ &  $266$ \\
        Twitter & $15,314$ & $666$  & $283$ & $4927$ \\
        Facebook & $1,146$ & $241$ &  $100$ & $684$ \\
        YouTube & -- & $667$ & $283$ & -- \\
                \hline
    \end{tabular}
    \label{tab:dataset}
\end{table}
\section{Methodology}\label{sec:methods}
%In this section, we introduce the different methods we use to detect and analyse coordinated information campaigns on social media platforms. 

%\ainote{need overview of the approach first}
Our objective is to design an approach based on content similarity and time locality that identifies groups of messages that are likely coordinated. 
Because we do not have ground truth information at fine granularity---that is, we do not know which are the particular messages or the user accounts involved in a campaign---we need to find different ways to evaluate the likelihood that our groupings make sense.

\begin{figure}
    \centering
    \includegraphics[scale=0.48]{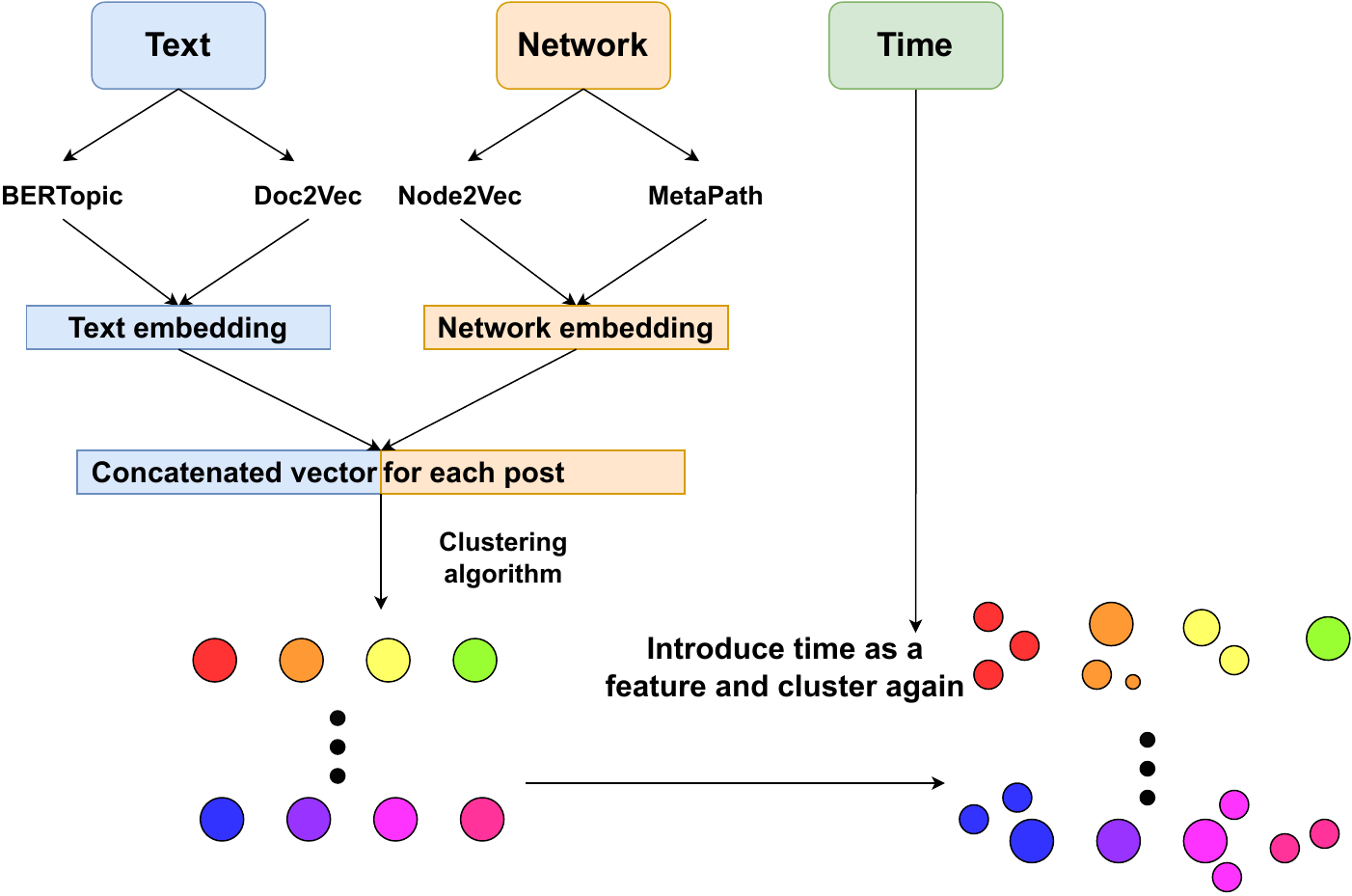}
    \caption{Methodology pipeline overview. Textual and network embeddings are concatenated, after which we perform a first clustering on the produced embeddings and a second clustering on the time component.}
    \label{fig:methods overview}
\end{figure}

The approach we propose makes some assumptions about coordinated information operations, listed in Section~\ref{sec:assumptions}. 
It includes two clustering stages, one based on text and network embeddings and one based on the time lapse between posts on social media platforms. 
We experiment with two text embeddings techniques, Doc2Vec and BERTopic, as presented in Section~\ref{sec:bertopic}.
Furthermore, we define a user-to-content network and extract its network embeddings using Node2Vec~\cite{grover_node2vec_2016} and, alternatively, MetaPath~\cite{dong_metapath2vec_2017}, that we append to our textual embeddings. 
Finally, we account for temporal dynamics by incorporating a final clustering component based on post time. 
The final pipeline leverages multimodal embeddings by combining text similarity, temporal information and the network structure of users, posts, and shared links. 
Figure~\ref{fig:methods overview} illustrates our processing pipeline.

%\jnote{playing with the details but i would rename the NLP part in the figure to text and the graph to network} \ainote{yes, that's important. Carlos?}

\subsection{Assumptions}
\label{sec:assumptions}

Our approach makes the following assumptions:
%\begin{itemize}
%    \item 
    
    \textit{Coordinated campaigns are characterized by messages posted closely in time and that discuss similar topics.} This may not always be the case: a coordinated campaign may publish posts throughout a longer period of time, and mostly exploit liking each other posts to make them more visible.
    
    %\item 
    
    \textit{Text embeddings of posts across social media platforms are comparable.} This implies that the language used across social media platforms is similar, and that the language model used will encode them similarly. In practice, this may not always be the case, for example, tweets are known to be much shorter than posts on Reddit.
    
    %\item 
    \textit{Links posted by coordinated campaigns are from sources with similar factuality scores.} The underlying idea is that a coordinated campaign is unlikely to be posting about sources with conflicting viewpoints. In practice, we might have reliable sources being discredited in a coordinated campaign, alongside other less reliable posts. 
    
    %\item 
%% \ainote{is this really different from anybody else? I might not understand it}    \textit{The distribution followed by the dataset used to train and evaluate the models is comparable to the one followed by the data in the entire platform.} %Limitations of this assumption are discussed in Section~\ref{sec:datasets}.
    
    %\item 
    Apart from the Node2Vec method, all other methods assume that \textit{past user's activity is not relevant for identifying coordinated campaigns.} This seems to differ among campaigns, and has been seen to be true in some campaigns, such as AIMS~\cite{ganguly_aims_2023}.
%\end{itemize}

%\jnote{Debating whether this paragraph (assumptions) needs to be here, or after the description of the methods. }

Given these assumptions, our methodology builds on temporal, network and textual proximity of posts with the goal of detecting potential coordinated campaigns. Having specific thresholds for the proximity allows us to identify groups (clusters) that are potentially part of the same campaign.
%Hence, by analysing the temporal proximity and semantic similarity of posts within each subcluster, we hope to detect potential coordinated campaigns in the dataset.
%This can be achieved by establishing thresholds for temporal proximity and semantic similarity and identifying clusters or subclusters that surpass these thresholds.

%\jnote{I restructured this a bit so that we have a section on assumptions and then one section for each part: text, network and time. Titles can change but i think it makes more sense this way.}

\subsection{Text (Semantic) Similarity}\label{sec:text}

We explored two approaches that can handle the text content of posts, namely Doc2Vec and BERTopic. They are both enhanced by a temporal component, as described in Section \ref{sec:time-exp}.

\subsubsection{Doc2Vec.}\label{sec:doc2vec_time}
Our first approach combines Doc2Vec with K-means. The following steps outline the process. We start with a Doc2Vec representation of each post. This technique enables us to effectively capture the semantic relationships between posts, facilitating subsequent clustering. We then perform an initial clustering with $K$-means to group semantically similar posts together. We determine the number of clusters based on the characteristics of the dataset.
%\jnote{research objectives doesn't sound good here}. 
To further refine the clustering results by accounting for temporal dynamics, we employ the Density-Based Spatial Clustering of Applications with Noise (DBSCAN) algorithm. We provide more details on the latter part in Section \ref{sec:time-exp}.

\subsubsection{BERTopic.}\label{sec:bertopic}
Our second approach involves BERTopic to obtain clusters for all posts with a topic representation of the clusters~\cite{grootendorst_bertopic_2022}. BERTopic uses Sentence-BERT to build 384 dimensional embeddings per document. We then apply UMAP for dimensionality reduction, mainly so the clustering is more efficient. For clustering, we employ HDBSCAN, a hierarchical density-based clustering algorithm. Lastly, we obtain our topic representations from clusters with c-TF-IDF, generating candidates by extracting class-specific words. To improve our topic representations, we end the pipeline with Maximum Candidate Relevance. Note, while we implement the same pipeline as the original paper, each component could be replaced by alternatives. For instance, one could employ PCA instead of UMAP for dimensionality reduction. As with Doc2Vec we apply DBSCAN to get clusters based on temporal information as well. 

\subsection{Network Similarity: Random Walks}\label{sec:random_walks}

\begin{figure}
    \centering
    %\includesvg[width=120pt]{graph_structure.svg}
    \includegraphics[width=180pt]{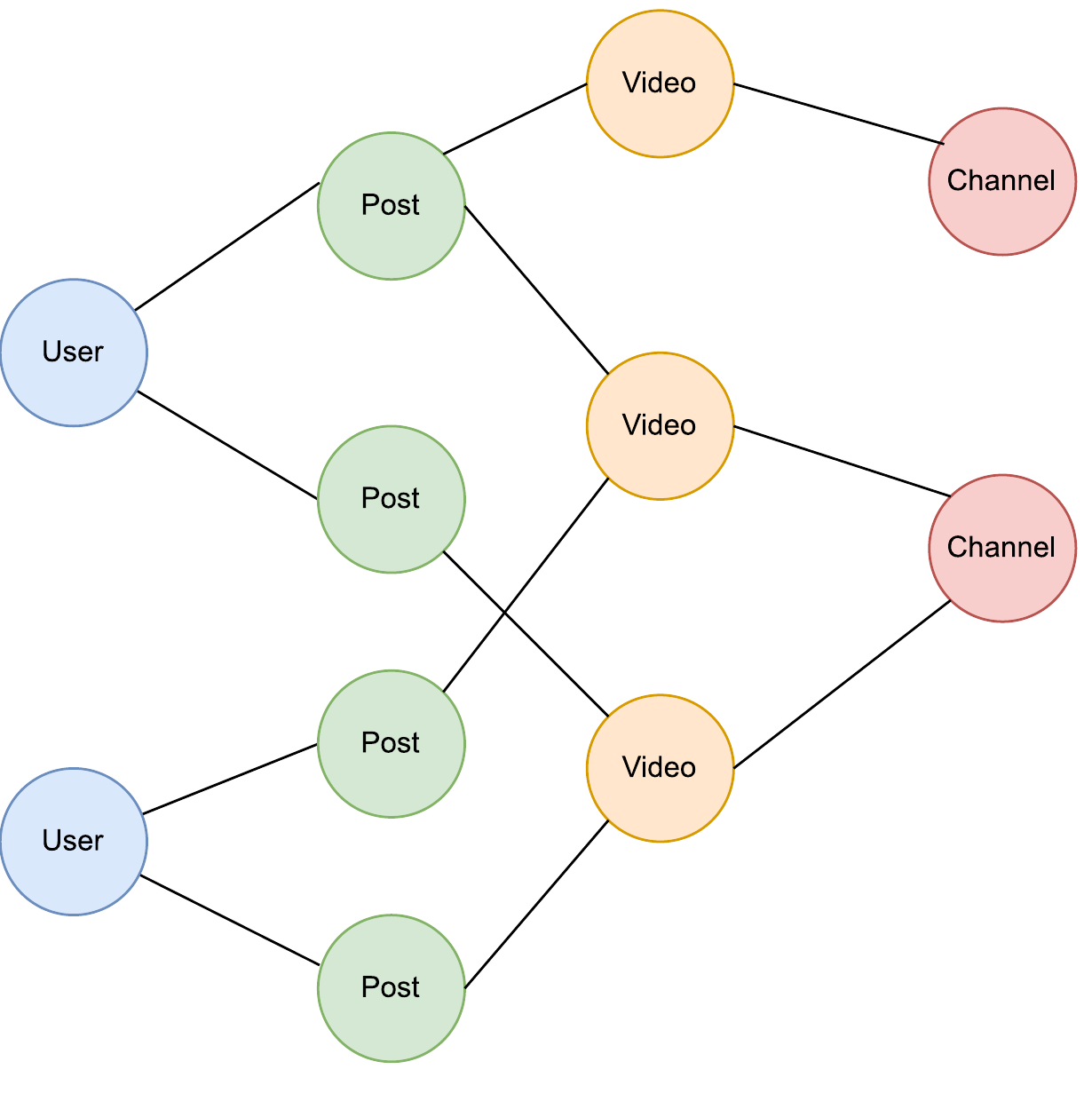}
    \caption{Example network structure.}
    \label{fig:graph-structure}
\end{figure}

We define a network that connects users across social media platforms as shown in Figure~\ref{fig:graph-structure}. 
In this multipartite network, edges connect users to their posts, posts to videos, and videos to channels. 
While this definition is specific to the dataset and social media platforms we used, it can be generalised to, for example, (URLS to) online articles (instead of YouTube videos) and their hosting websites (instead of YouTube channels). 
%This network is much smaller, as not all nodes are connected. \ainote{much smaller than what? you mean, disconnected?}
This structure does not capture text or temporal information but only information on user interaction with content (URLs). However, this allows us to create graph embeddings for each post using Node2Vec, as nodes are only sparsely connected. 
We concatenate these network embeddings to the text embeddings described in Section~\ref{sec:text}. 
To align both embeddings, we perform dimensionality reduction on both representations, such that both embedding vectors have the same length ($64$ in our case).
%\jnote{This can be debated by reviewers: why 64 and why did we do further DR here.} \knote{DR was just to match both embeddings in size. There isn't a particular reason to why 64 to be honest.}

Node2Vec does not take into account the different typed of nodes in a network. 
In our network, there are four types of nodes: users, posts, videos and channels. 
Since this information might be important and to account for these different types of nodes, we apply meta-paths~\cite{dong_metapath2vec_2017}. 
Meta-paths guide the random walks by only allowing a fixed set of paths. In this paper, we set the following paths: [user, post, video, post, user]; [video, channel, video]; [post, user, post] and [channel, video, channel]. This allows us to incorporate semantic information on node types into the network embeddings.

\subsection{Temporal Dynamics}\label{sec:time-exp}
We assume coordinated campaigns occur within a short timeframe. By considering the temporal dimension, we ensure that we detect subgroups (subclusters) containing posts that are not only semantically similar, but also close in time (i.e. posted close in time). Hence, we subdivide clusters obtained from Doc2Vec and BERTopic to account for temporal dynamics through a DBSCAN algorithm. We choose $\epsilon=52$ seconds as suggested by~\cite{ng_multi-platform_2021} for the same dataset we used. This is the inter-arrival time threshold in seconds between a video's first share on any platform (either Twitter or Facebook) and all successive shares. For more details on our choice of $\epsilon$, we refer to~\cite{ng_multi-platform_2021}. With this approach accounting for temporal dynamics, we aim to facilitate the discovery of underlying patterns and trends, shedding light on coordinated activities that may impact the analysed content and enhancing our understanding of the data and identifying potential coordinated efforts.
\section{Results}\label{sec:results}

In this section, we first describe the metrics we use to evaluate our methods. We then perform an ablation study among the three components (text, network and time) of our approach to evaluate how each affects the performance of our approach. Finally, we further investigate different values for certain hyperparameters.

\subsection{Metrics}
To evaluate the effectiveness of our approach, we employ two metrics: the standard deviation of factuality scores among clusters and the silhouette score for the clustering of Doc2Vec, BERTopic and Random Walks features before and after temporal subdivision. The following analysis provides insights into the performance and detection capabilities of our approach.

\textbf{Silhouette Score:}
We assess the quality of the clustering results by computing the silhouette score~\cite{rousseeuw_silhouette_1987}. The silhouette score measures the cohesion and separation of data points within clusters. A higher silhouette score indicates better-defined and more distinct clusters. We compute the silhouette score using the features before and after applying the temporal subdivision, providing insights into the improvement achieved through the incorporation of temporal considerations.

\subsubsection{Standard Deviation of Factuality Scores}
We calculate the factuality score for posts containing YouTube links, which rates the factual accuracy of the information in the respective YouTube channels. After executing the entire pipeline, we analyse the factuality score for each cluster. A low standard deviation among the factuality scores indicates that our algorithms cluster posts sharing videos of similar factuality, whether they are questionable or of high factuality, whereas a higher standard deviation suggests potential variations in the factual accuracy of the information within the clusters. More specifically, we report the average, median, standard deviation, and proportion of $0$ standard deviation in clusters among all clusters. As a comparison, we provide the standard deviation of all posts.

\subsection{Evaluation: Ablation Study}

In this section, we discuss the performance of the different methods. To evaluate the performance of our clustering, we compare the silhouette score and the intra-cluster standard deviation in factuality score. As mentioned in our assumptions in Section~\ref{sec:assumptions}, we expect coordinated campaigns on social media to share sources of similar factuality, whether questionable or high factuality. As we end the pipelines of both methods by subclustering with regards to the posts' time, we compare the factuality standard deviation pre- and post-time clustering, and the effect of our graph embeddings on these metrics. Our evaluation results can be found in Table~\ref{tab:bertopic_doc2vec}.

%\ainote{Does this mean that when we do not use time we also include the clusters of 1 item only?} \knote{we explained this wrongly. This is for both post time and pre time. I corrected it.}
\begin{table}[htbp]
    \centering
    \caption{Evaluation of the different approaches. We only keep clusters of size of at least $2$ posts to evaluate the factuality standard deviation. $\propto 0$ denotes the proportion of Factuality Standard Deviation of $0$ obtained through all clusters. The total standard deviation in the dataset is of $0.75$. N2V and MP2V denote Node2Vec and Metapath2Vec, respectively.} 
    \begin{tabular}{l|ccccc}
        \hline
        \multirow{2}{*}{Methods} & \multirow{2}{*}{\shortstack{Silhouette\\Score}} & \multicolumn{4}{c}{Factuality Standard Deviation} \\ \cline{3-6}
        &  & \makebox[1.1cm]{Avg} & \makebox[1.1cm]{Median} & \makebox[1.1cm]{Std} & \makebox[1.1cm]{$\propto 0$} \\ \hline
         Doc2Vec & $0.04$ & $0.61$ & $0.65$ & $0.49$ & 0.26 \\
         Doc2Vec+PostTime & $-0.17$ & $0.38$ & $0.26$ & $0.41$ & 0.10\\
         BERTopic& $0.30$ & $0.32$ & $0.18$ & $0.41$ & $0.24$\\
        BERTopic+PostTime & $-0.26$ & $\mathbf{0.02}$ & $\mathbf{0.00}$ & $0.13$ & $\mathbf{0.96}$\\
         Doc2Vec+N2V& $0.23$ & $0.52$ & $0.41$ & $0.48$ &0.05\\
         Doc2Vec+PostTime+N2V& $-0.07$ & $0.13$ & $1.45$ & $0.27$ &0.33\\
         BERTopic+N2V& $\mathbf{0.54}$ & $0.09$ & $\mathbf{0.00}$ & $0.27$ & $0.54$\\
         BERTopic+PostTime+N2V& $-0.61$ & $\mathbf{0.02}$ & $\mathbf{0.00}$ & $\mathbf{0.12}$ & $\mathbf{0.96}$\\
         Doc2Vec+MP2V& $0.23$ & $0.56$ & $0.57$ & $0.43$ &0.2\\
         Doc2Vec+PostTime+MP2V& $-0.11$ & $0.19$ & $2.79$ & $0.31$ &0.16\\
         BERTopic+MP2V& $0.31$ & $0.27$ & $0.00$ & $0.40$ & $0.37$\\
         BERTopic+PostTime+MP2V& $-0.44$ & $\mathbf{0.02}$ & $\mathbf{0.00}$ &$0.15$ & $0.94$\\
        % BERTopic + Post Time

         %BERTopic & &&&&&\\
         \hline
    \end{tabular}
    \label{tab:bertopic_doc2vec}
\end{table}

The silhouette score drops after clustering the time component for both methods. This is because we obtain clusters of significantly smaller size. Post-time subclustering is computed on the time component, and not based on the original embeddings, while the silhouette score is still computed with respect to the original embeddings. Due to the temporal subclustering, the new clusters are no longer optimal in respect to the original embeddings (which are taking into account only the text content). BERTopic encapsulates the textual information better than Doc2Vec according to the silhouette score of the two methods. Still, the silhouette score is far from $1$, the optimal value. Values close to 0 signify clusters that overlap. Negative results typically signify that a post was attributed to a wrong cluster because another cluster would have better fit the post.

All methods achieve a better average factuality standard deviation than the total standard deviation in the dataset. Clustering post-time strongly improves the factuality standard deviation. This highlights the temporal aspect plays a role in coordination campaigns, as this improves the clusters' factuality wise.

BERTopic+PostTime identifies $5,544$ clusters of only one post. We drop those clusters for our analysis as we judge they are unlikely to be part of a coordinated campaign. Hence, we suspect the $587$ clusters of size at least $2$ ($\approx 10\%$ of the clusters) to be likely part of a coordinated campaign as they have similar textual and temporal information (the posts within a cluster have been shared within $52$ seconds apart). Interestingly, the factuality standard deviation strongly reduces to an average of $0.01$ post time ($\approx 97\%$ lower than pre time), with $98\%$ of the clusters having a factuality score standard deviation of $0$.

Overall, BERTopic+PostTime+N2V performs best in factuality standard deviation. BERTopic produces contextual embeddings that are taking into account the context of the document as opposed to Doc2Vec, a simple generalization of Word2Vec. In most cases, incorporating network embeddings improves the performance, with Node2Vec scoring better. An explanation for the Node2Vec performing better than MetaPath2Vec could be that we bias random walks in Node2Vec to be more likely to be directed at nodes that the walk has already visited. We do this because coordinated posts usually reference the same or similar videos. These nodes will, therefore, be close together in the graph. Biasing the random walks to explore less puts more focus on these local subgraphs, which gives posts close together in this graph more similar embeddings. The exact values we used were $p=0.25$ and $q=4$. In the implementation of the meta paths we use, these cannot be set.

Not included here due to space constraints, we also considered different values for the hyperparameters $K$ and $\epsilon$. 
$K$ is used in the $K$-means algorithm. 
$\epsilon$ is the radius of the DBSCAN search algorithm, in seconds. 
Additionally, we consider two cases Twitter, Facebook as the first case and the Reddit dataset as the second case.
We further investigate the importance of choosing the right $\epsilon$ threshold. 
We evaluate the factuality standard deviation for different value of $\epsilon$. % in Appendix \ref{app:temporal} (Table \ref{tab:time_epsilon} and Figure \ref{fig:silhouette_per_epsilon}). 
The optimal values with regard to the factuality standard deviation seem to lie in the interval $[30,60]$ seconds, confirming our choice of $52$ seconds remains close to the optimal choice of $\epsilon$.
%\section{Discussion}\label{sec:discussion}
\subsection{Coordinated Campaigns}
BERTopic+PostTime+MP2V performs best in factuality standard deviation. The clusters obtained by this approach take textual information into account, through Sentence-BERT embeddings. Furthermore, temporal aspect is incorporated with our clustering where each post within a cluster must have another post in that cluster that was posted within $52$ seconds. Lastly, it encompasses the social media interaction through the Node2Vec embeddings of our graph. Since our clusters incorporate all those aspects, and due to the strong factuality standard deviation scores, we believe these posts are highly likely to be part of a coordinated campaign. In the remaining of this section, we perform an analysis of these clusters.
\begin{figure}
  \centering
  \begin{subcaptiongroup}
    \centering
    \parbox[b]{.5\textwidth}{%
    \centering
    \includegraphics[width=0.5\textwidth]{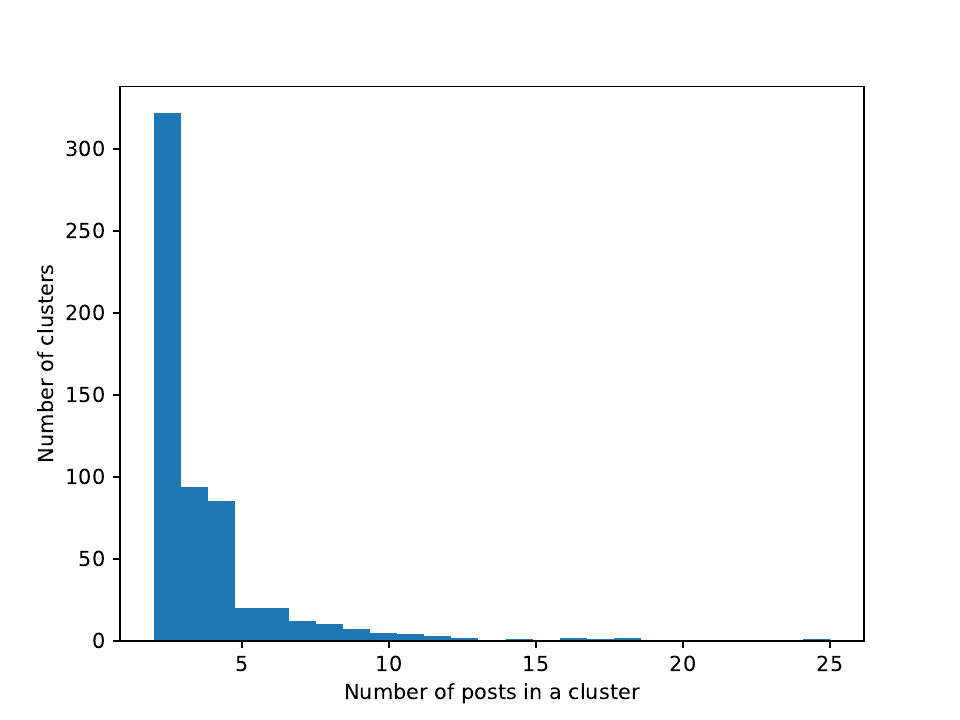}
    \caption{Distribution of posts per cluster.}\label{fig:posts_per_cluster}}%
    \parbox[b]{.5\textwidth}{%
    \centering
    \includegraphics[width=0.5\textwidth]{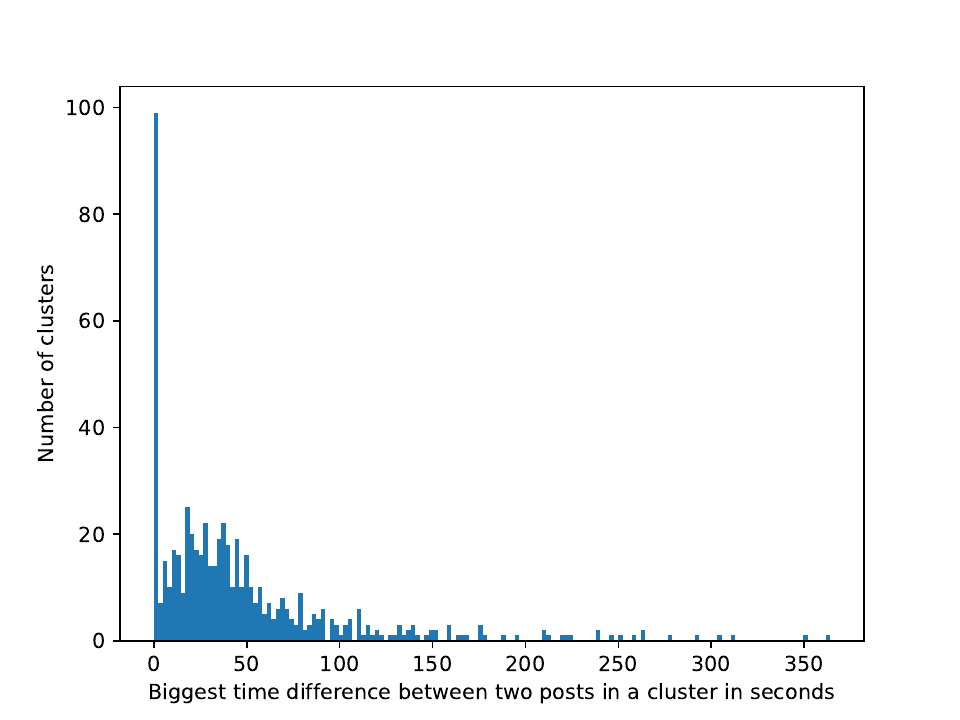}
    \caption{Time difference between two posts.}\label{fig:timediff_per_cluster}}%
  \end{subcaptiongroup}
  \caption{Distribution of posts (Figure \ref{fig:posts_per_cluster}) and largest time difference between two posts within a cluster (Figure \ref{fig:timediff_per_cluster}) for clusters obtained by BERTopic+PostTime+N2V.}\label{fig:cluster_analysis}
\end{figure}

 BERTopic+PostTime+MP2V outputs $591$ clusters with $2098$ posts in total ($1920$ tweets, $141$ Facebook posts and $37$ Reddit comments), which represents $8\%$ of the total posts in our dataset, hence an average of $3.37$ posts per cluster. Figure \ref{fig:posts_per_cluster} shows the distribution of posts per cluster. The distribution is right skewed, with a maximum of $24$ posts within a cluster, despite a significant number of clusters with only two posts. Despite a low $\epsilon$ of $52$ seconds in our DBSCAN algorithm, some clusters have posts separated by up to six minutes (Figure \ref{fig:timediff_per_cluster}). This is still a very short timeframe, particularly considering the time span of our dataset over a year. As a matter of fact, most of the posts within a cluster are separated by $100$ seconds or less. This strong time similitude highlights a coordination pattern between posts. 
 %Examples of clusters are shown in Appendix \ref{app:example-clusters}, including an example of a cross-platform cluster.
 Most of the clusters have posts with the exact same textual information and a different post time. We observe that $89\%$ of the channels embedded in the posts had a factuality score of $0$, hinting most coordinated behaviour covers questionable sources of very low factuality. We show the $10$ most frequent video channels in Table~\ref{tab:most_frequent_video_channels}.

 \begin{table}[htbp]
    \centering
    \caption{The $10$ most frequent video channels in the clusters computed by BERTopic+PostTime+MP2V. The description provided below was not used in our framework for identifying coordinated information operations, and is from online sources shown in parenthesis and collected in July 2023.} 
  %  \begin{tabular}{l|r|p}
 \begin{tabularx}{\textwidth}{l|r|p{6.5cm}}
    
    \hline
        Video channel & \# Posts & Description\\
        \hline
        Clarity Of Signal & $1,204$ & Focused on Syrian war, 8 videos and 200 subscribers [YouTube description].\\
        \hline
        Vanessa Beeley & $455$ & British activist and blogger known for sharing conspiracy theories and disinformation about the Syrian civil war and the White Helmets.[Wikipedia]\\
        \hline
        RT&$244$ & ``Questionable based on promoting pro-Russian propaganda, conspiracy theories, numerous failed fact checks, and a lack of author transparency.'' [MBFC]\\
        \hline
        Corbett Report&$76$ & ``Overall, we rate the Corbett Report a Tin Foil Hat conspiracy and Moderate pseudoscience website, based on the promotion of 9/11 conspiracies, False Flags, Chemtrails, and Deep State conspiracies.'' [MBFC] \\
        \hline
        RT UK&$20$ & (see RT)\\
        \hline
        Corbett Report Extras&$16$ & ``This is the secondary channel of The Corbett Report.'' [YouTube description]\\
        \hline
        Syriana Analysis&$14$ & ``Overall, Syriana Analysis has a pro-Assad view (The Arab Socialist Bath Party – Syria Region); therefore, we rate Syriana Analysis Left biased due to their story selection and pro-Assad view. We also rate them Questionable due to poor sourcing, opinion-based commentary, lack of transparency, and a failed fact check'' [MBFC]\\
        \hline
        The Last American Vagabond&$12$ & Meanwhile deleted by YouTube. \\ 
        \hline
        Sputnik& $11$ & ``Questionable based on the frequent promotion of conspiracies and pro-Russian propaganda, as well as the use of poor sources and numerous failed fact checks.'' [MBFC]\\
        \hline
        RT America &$11$ & (see RT)\\
        \hline
    \end{tabularx}
    \label{tab:most_frequent_video_channels}
\end{table}

\subsection{Limitations}
The limitations of this project can be categorized into two main areas: limitations related to the underlying models' structure and limitations of the results obtained from training the models on the given data.

%\subsubsection{Modelling and Ideation}
Working with social media data is known to be particularly tricky, and this project is no exception.
In particular, the patterns that our methods are trying to clusters are the result of controlled human behaviour, meaning that if these methods were put into place to identify clusters in an online scenario they would be working against "attackers" that will try to fool the system.
Our solution would need a dynamically inferred time threshold like the one used to infer the 52 seconds interval in previous work~\cite{ng_multi-platform_2021}.  

% \ainote{not sure} One significant limitation of the models used in this study is their black-box nature, as they lack transparency in explaining why a particular post is assigned to a specific cluster. Consequently, these models should only serve as post-hoc analysis tools, providing insights into potential patterns within the data. It is crucial not to employ such models in real-time online scenarios, as they are computationally expensive and incapable of providing conclusive results.

%\subsubsection{Data and results}
A limitation of the results stems from the assumption that embeddings across different social media platforms can be directly compared within the same model, without considering the platform-specific language and behaviour. However, it is evident that the language used on Twitter greatly differs from that used on Facebook or Reddit, meaning the evaluation of our model acts as a lower bound. Furthermore, attackers are likely to employ different strategies tailored to each social media platform to exploit each social media's algorithm effectively.
%The data provided to us contained some inconsistencies, such as the YouTube Video ID only having been retrieved at a later stage, making some results unreliable. This means that two different posts linking to the same video in our dataset can potentially be talking about different topics in reality.

%The data used in this study focused solely on one topic, which limits the generalizability of the findings.

As discussed in Section~\ref{sec:datasets}, there is no ground truth of what posts belong to coordinated campaigns. 
Our approach to use Media Bias Fact Check for evaluation has two limitations: one one hand, not all channels are evaluated; on the other hand, the evaluation is at the news site granularity and not individual content (video, in our case). 
Thus, it is theoretically possible that a particular video in our dataset was fact-based while being hosted on a low factuality YouTube channel. 
That is why qualitative analysis of the content flagged by our approach is recommended.
%, in practice su%is limited: a particular video marked as "unreliable" might actually be unbiased reporting, while the source behind the channel might be labeled as low factuality in its general reporting.
% \section{Future Work}
% we don't have future work so i think just a sentence in the conclusion should be enough

\section{Summary}
This paper proposes an approach for identifying likely coordinated messages on multiple social media platforms that integrates different modalities in addition to temporal dynamics. 
We tested our approach on a dataset that includes content posted on Twitter, Facebook, Reddit, and YouTube related to the White Helmets, the Syrian Civil Defence organization that was shown to be the target of discrediting information operations.
Our method incorporates three key modalities relevant for a coordination campaign: i)~textual information (through Doc2Vec and BERTopic); ii)~temporal information, to detect social media posts with good temporal locality; and iii)~content and user interactions represented as Node2Vec embeddings of a multi-modal network of users, posts, YouTube videos, and YouTube channels. 

%We encountered and discussed several challenges in working with social media datasets, including the lack of labelled data, difficulties in obtaining representative and comprehensive data, and limited access to metadata. 

To evaluate the effectiveness of our approach in the absence of ground truth, we used metrics such as the standard deviation of factuality scores and the silhouette score. 
The standard deviation of factuality scores allowed us to assess the consistency of factual accuracy within clusters, while the silhouette score measured the quality of clustering results before and after incorporating temporal considerations. 
The results strongly hint we managed to find clusters of posts that are suspiciously coordinated in time and low-credibility content.

In the future we plan to test our approach on other multi-platform social media datasets, some that are likely to contain traces of coordinated information operations and others that do not. 
While in this paper we focused on designing an approach that includes the necessary elements of a multi-platform coordinated campaign---locality in content and in time---we plan to further evaluate the accuracy of our identification approach, and its adaptability to different platforms.  %e are interested nethat are both likely and unlikely to 

% \newpage
\bibliographystyle{splncs04}
\bibliography{references}

\end{document}